\documentclass[preprint,showpacs,prd,amsmath,amssymb]{revtex4}
\usepackage{graphicx}

\renewcommand{\bar}[1]{\overline{#1}}

\begin{document}

\preprint{USM-TH-135}

\title{Mass Suppression in Octet Baryon Production}

\author{Jian-Jun Yang}
\email{Jian-Jun.Yang@physik.uni-regensburg.de}
\affiliation{Department of Physics, Nanjing Normal
University, Nanjing 210097, China}
\affiliation{Institut f\"ur Theoretische Physik,
Universit\"at Regensburg, D-93040 Regensburg, Germany}

\author{Ivan Schmidt}
\email{Ivan.Schmidt@fis.utfsm.cl}
\affiliation{Departamento de F\'\i sica, Universidad
T\'ecnica Federico Santa Mar\'\i a, Casilla 110-V, Valpara\'\i so, Chile}

\begin{abstract}

There is a striking suppression of the cross section for
production of octet baryons in $e^+ e ^-$ annihilation, as the
mass of the produced hadron increases. We present a simple
parametrization for the fragmentation functions into octet baryons
guided by two input models: the SU(3) flavor symmetry part is
given by a quark-diquark model, and the baryon mass suppression
part is inspired by the string model. We need only {\it eight} free
parameters to describe the fragmentation functions for all octet
baryons. These free parameters are determined by a fit to the
experimental data of octet baryon production in $e^+ e ^-$
annihilation. Then we apply the obtained fragmentation functions
to predict the cross section of the octet baryon production in
charged lepton DIS and find consistency with the available
experimental data. Furthermore, baryon production in $pp$
collisions is suggested to be an ideal domain to check the
predicted mass suppression.

\end{abstract}


\maketitle

\section{Introduction}

It is well known that there is an SU(3) flavor symmetry between
the wave functions of the octet baryons. Very recent lattice QCD
calculations also suggest that the $\Lambda$ and proton quark
structures can be related by an SU(3)
transformation~\cite{Lattice}. Unfortunately it is not possible to
investigate the parton distributions of the octet hyperons by
means of structure functions in deep inelastic scattering (DIS),
since they cannot be used as a target due to their short
life-time. Also one obviously cannot produce a beam of
charge-neutral hyperons such as $\Lambda$. Therefore there is no
experimental information on the relation between the parton
distributions of octet hyperons and those of the nucleon. However,
it has been proposed that the quark fragmentation functions might
be related to the corresponding quark distributions by a so called
Gribov-Lipatov relation~\cite{GLR,Bro97}. Therefore we can explore
the relation between quark distributions of octet baryons by means
of hyperon production from quark fragmentation. Actually great
progress has been made along this direction
recently~\cite{MSSY,MSY}.

A specific regularity in baryon production rates in $e^+e^-$
annihilation has been recently noticed by Chliapnikov and
Uvarov~\cite{CU95}. However, to our knowledge, there has not been
a detailed investigation on the relation among the octet baryon
fragmentation functions. The data shows a striking suppression of
the cross section as the mass of the produced hadron increases,
which is puzzling since naive SU(3) would suggest that they should
be comparable. For example, once an s-quark is produced then one
could think that the $\Sigma :\Lambda$ ratio would be $3:1$, since
the $qq$ pair is statistically three times more likely to be in an
isospin $I=1$ state than in isospin $I=0$. Nevertheless the
experimental data shows just the opposite trend, with a lower
cross section for $\Sigma$ production (see Fig. \ref{ys1f1}).
Apparently something in the fragmentation process makes it much
easier for the s-quark to pick up an isosinglet $qq$ than an
isotriplet.

In this paper we will not provide a detailed physical explanation
of the suppression. Our, more modest, purpose is to give a
consistent simple parameterization of all the fragmentation
functions into octet baryons, with a rather small number of free
parameters. Nevertheless, since the parameterizations that we
obtain are inspired by well known input models, they can be
considered a first step into a physical understanding of these
processes. For every octet baryon, there are 18 (unpolarized,
longitudinally polarized and transversely polarized)
quark/antiquark fragmentation functions. As a sensible
parametrization of a fragmentation function usually needs at least
3 parameters, a lot of experimental data are needed in order to
fix all of them. We plan to constrain the shape of fragmentation
functions with the help of some models in order to reduce the
number of free parameters. Models such as those using strings and
shower algorithms~\cite{Barger} still involve many parameters, and
therefore with these models it is difficult to obtain a clear
relation among the fragmentation functions for various octet
baryons. On the other hand, the diquark model given in
Ref.~\cite{Nza95} can provide us with SU(3) flavor symmetry
relations for octet baryon fragmentation functions. An
investigation of the spin structure of the diquark fragmentation
functions can be found in previous
publications~\cite{Nza95,a01,a06}. It was found that the spin
structure of the diquark model fragmentation functions for the
$\Lambda$ is supported by the available experimental
data~\cite{a01,a06}, which indicates that the diquark model works
well in describing the fragmentation functions. In the present
work, we will retain the flavor and spin structure of
fragmentation functions as given by the diquark  model and
concentrate on the relation between fragmentation functions of
various octet baryons. We will emphasize the mass suppression
effect in hyperon production.

The paper is organized as follows. In Sec.~II, we briefly describe
our ansatz of the fragmentation functions for all octet baryons
based on the diquark model and the string model. Using only eight
free parameters, we relate the fragmentation functions for all
octet baryons to each other. The free parameters of the model are
determined based on the experimental data on octet baryons
production in $e^+e^-$ annihilation, where the mass suppression
effect in hyperon production is very important in order to
understand the available experimental data. In Sec.~III, we
propose a possible cross check of the mass suppression via  octet
baryon production in charged lepton DIS. In Sec.~IV, we give
predictions for cross sections of baryon production in $pp$
collisions with the obtained fragmentation functions for the octet
baryons. Finally, we give a summary with our conclusions in
Sec.~V.

\section{Quark fragmentation functions for the octet baryons}

Recently, it was found that a simple diquark model can be used to
describe quite accurately the octet baryon fragmentation
functions~\cite{Nza95,a01,a06}. The parameters of the model were
determined by fitting the experimental data of octet baryon
production in $e^+e^-$ annihilation. In this work, we adopt an
alternative approach and present a parametrization based on the
spin and flavor structure predicted by the diquark model. We focus
our attention on the relation between the fragmentation functions
of various octet baryons.

In contrast to the nucleon parton distributions which are well
determined by  experimental data, we have much less information on
fragmentation functions for the octet baryons. For this reason, we
constrain the fragmentation functions with the help of some models
in order to reduce the number of free parameters. In particular
the diquark model~\cite{Nza95} has a clear physical motivation and
needs only a few parameters.

Within the framework of the diquark model~\cite{Nza95}, the
unpolarized valence quark to  proton fragmentation functions can
be expressed as

\begin{equation}
D_{u_v}^p(z)= \frac{1}{2} a_S^{(u/p)}(z)
+ \frac{1}{6} a_V^{(u/p)}(z),
\end{equation}

\begin{equation}
D_{d_v}^p(z)= \frac{1}{3} a_V^{(d/p)}(z),
\end{equation}
where $a_D^{(q/p)}(z)$  ($D=S$ or $V$) is the probability of
finding a quark $q$ splitting into the proton $p$ with
longitudinal momentum fraction $z$ and emitting a scalar ($S$) or
axial vector ($V$) anti-diquark. The form factors for scalar and
axial vector diquark are customarily taken to have the same form

\begin{equation}
\phi (k^2)=N \frac{k^2-m^2_q}{(k^2- \Lambda_0^2)^2},\label{form}
\end{equation}
with a normalization constant $N$ and a mass parameter
$\Lambda_0$. The value $\Lambda_0=500~\rm{MeV}$ is usually adopted
in numerical calculations. In (\ref{form}), $m_q$ and $k$ are the
mass and  the momentum of the fragmenting quark $q$, respectively.
According to Ref.~\cite{Nza95}, in the quark-diquark model
$a_D^{(q/p)}(z)$ can be expressed as

\begin{equation}
a_D^{(q/p)}(z)=\frac{N^2 z^2 (1-z)^3}{64 \pi^2}
\frac {[2 ( M_p + m_q z)^2+
 R^2(z)]}{R^6(z)}\label{ad}
\end{equation}
with

\begin{equation}
R(z)=\sqrt{z m_D^2-z(1-z)\Lambda_0^2+(1-z)M_p^2},
\end{equation}
where $M_p$ and $m_D$~($D=S$ or $V$) are the mass of the proton
and diquark, respectively. We choose the values for the diquark
masses to be $m_S=900~ \rm{MeV}$ and $m_V=1100~ \rm{MeV}$, for
scalar and axial vector diquark states, respectively. The quark
masses are taken as $m_u=m_d= 350~ \rm{MeV}$.

Similarly, the longitudinally and transversely  polarized quark to
proton  fragmentation functions can be written  as

\begin{equation}
\Delta D_{u_v}^p(z)
= \frac{1}{2} \tilde{a}_S^{(u/p)}(z)
- \frac{1}{18} \tilde{a}_V^{(u/p)}(z),
\end{equation}

\begin{equation}
\Delta D_{d_v}^p(z)= -\frac{1}{9} \tilde{a}_V^{(d/p)}(z),
\label{Dsv}
\end{equation}

\begin{equation}
\delta D_{u_v}^p(z)
= \frac{1}{2} \hat{a}_S^{(u/p)}(z)
- \frac{1}{18} \hat{a}_V^{(u/p)}(z),
\end{equation}

\begin{equation}
\delta D_{d_v}^p(z)= -\frac{1}{9} \hat{a}_V^{(d/p)}(z),
\label{dsv}
\end{equation}
with

\begin{equation}
\tilde{a}_D^{(q/p)}(z)=\frac{N^2 z^2 (1-z)^3}{64 \pi^2}
\frac {[2( M_p + m_q z)^2-R^2(z)]}{R^6(z)},\label{adb}
\end{equation}
and

\begin{equation}
\hat{a}_D^{(q/p)}(z)=\frac{N^2 z^2 (1-z)^3}{32 \pi^2}
\frac {( M_p + m_q z)^2}{R^6(z)},\label{adt}
\end{equation}
for $D=S$ or $V$. Here we are not interested in the absolute
magnitude of the fragmentation functions but in the flavor and
spin structure of them, which is given by the diquark model. In
order to extract the flavor and spin structure information, we
introduce flavor structure ratios

\begin{equation}
F_V^{(u/d)}(z)=\frac{a_V^{(u/p)}(z)}{a_V^{(d/p)}(z)},
\end{equation}

\begin{equation}
F_M^{(u/d)}(z)=\frac{a_S^{(u/p)}(z)}{a_V^{(d/p)}(z)},
\end{equation}
and spin structure ratios

\begin{equation}
W_D^{(q/p)}(z)=\frac{\tilde{a}_D^{(q/p)}(z)}{a_D^{(q/p)}(z)},
\end{equation}

\begin{equation}
\hat{W}_D^{(q/p)}(z)=\frac{\hat{a}_D^{(q/p)}(z)}{a_D^{(q/p)}(z)},
\end{equation}
with $D=S$ or  $V$.
Then we can use the fragmentation function $D_{d_v}^p (z)$ to
express all other unpolarized and polarized fragmentation functions
for the proton as follows

\begin{equation}
D_{u_v}^p (z)=\frac{1}{2}[ F_V^{(u/d)}(z) + 3 F_M^{(u/d)}(z) ]
D_{d_v}^p (z), \label{eq16}
\end{equation}

\begin{equation}
\Delta D_{u_v}^p (z)=\frac 32 [ W_S^{(u/p)}(z)  F_M^{(u/d)}(z) -
\frac{1}{9} W_V^{(u/p)}(z)  F_V^{(u/d)}(z) ]
D_{d_v}^p (z),
\end{equation}

\begin{equation}
\Delta D_{d_v}^p (z)= -\frac{1}{3} W_V^{(d/p)}(z) D_{d_v}^p (z),
\end{equation}

\begin{equation}
\delta D_{u_v}^p (z)=\frac 32 [ \hat{W}_S^{(u/p)}(z)  F_M^{(u/d)}(z) -
\frac{1}{9} \hat{W}_V^{(u/p)}(z)  F_V^{(u/d)}(z) ]
D_{d_v}^p (z),
\end{equation}
and

\begin{equation}
\delta D_{d_v}^p (z)= -\frac{1}{3} \hat{W}_V^{(d/p)}(z) D_{d_v}^p (z).
\label{eq20}
\end{equation}

The spin structure of the quark-diquark fragmentation functions
for the $\Lambda$ has been studied before~\cite{a01,a06}, and it
is supported by the available experimental data on $\Lambda$
production in various processes~\cite{ALEPH96, DELPHI95, OPAL97,
HERMES, E665, NOMAD}. In this work, we retain the flavor and spin
structure of the fragmentation functions suggested by the diquark
model.

From the above analysis, we find out that the essential ingredient
is to choose a suitable shape for the function $D^p_{d_v}(z)$ from
which the other valence fragmentation functions can then be
deduced. We could use the expression of $D^p_{d_v}(z)$ coming from
the diquark approach and  fix the parameters of the model by a fit
to the experimental data as it was done in Ref.~\cite{a06}.
Another way is to use a commonly accepted parametrization form
such as

\begin{equation}
D_{d_v}^p (z, Q^2_0) = N_v z ^{\alpha_v} (1-z) ^{\beta_v}, \label{fit1}
\end{equation}
with the exponents $\alpha_v$ and $\beta_v$ at an initial scale
$Q^2_0$. Previous work~\cite{a01,a06} indicates  that compatible
results can be obtained in both ways. In our present analysis, we
adopt the later approach, with the analytical expression
(\ref{fit1}) for $D^p_{d_v}(z)$, since this simple parametrization
can be easily used later for other purposes. In addition, the
diquark model fragmentation functions are easier to describe in
the large $z$ region where the valence quark contribution
dominates. In the small $z$ region, the sea contribution is
difficult to include in the framework of the diquark model.
Nevertheless, we also adopt a similar functional form

\begin{equation}
D_{q_s}^p (z, Q^2_0) = D_{\bar{q}}^p (z, Q^2_0)= N_s z ^{\alpha_s}
(1-z) ^{\beta_s} \label{fit2}
\end{equation}
to parameterize fragmentation functions of the  sea quark
$D_{q_s}^p (z)$ and antiquark $D_{\bar{q}}^p (z)$ for $q=u, d, s$
at the initial scale $Q^2_0$. For simplicity, we take the same
initial parametrization for the spin independent gluon and the sea
quark fragmentation functions, and moreover we assume that $\Delta
D_g^p$, $\delta D_g^p$, $\Delta D_{q_s(\bar{q})}^p$, and $\delta
D_{q_s(\bar{q})}^p$ at the initial scale are zero and that they
are only generated by QCD evolution.

Hence, the input unpolarized and polarized quark to proton
fragmentation functions can be written as

\begin{equation}
D_{q}^{p[\rm{SU}(3)]} (z, Q^2_0) = D_{q_v}^p (z, Q^2_0) + D_{q_s}^p (z, Q^2_0),
\end{equation}

\begin{equation}
\Delta D_{q}^{p[\rm{SU}(3)]} (z, Q^2_0) = \Delta D_{q_v}^p (z, Q^2_0),
\end{equation}
and

\begin{equation}
\delta D_{q}^{p[\rm{SU}(3)]} (z, Q^2_0) = \delta D_{q_v}^p (z, Q^2_0).
\end{equation}

Now we deduce the fragmentation functions
$D_{q(g)}^{B[\rm{SU}(3)]}$ for all other octet baryons $B$ by
SU(3) symmetry  at the initial scale $Q^2_0$. More specifically,
we have

\begin{equation}
\begin{array}{lllc}
D_u^{p[\rm{SU}(3)]}=D_d^{n[\rm{SU}(3)]}=
D_u^{{\Sigma^+}[\rm{SU}(3)]}=D_d^{{\Sigma^-}[\rm{SU}(3)]}
=D_s^{{\Xi^-}[\rm{SU}(3)]}=D_s^{{\Xi^0}[\rm{SU}(3)]}\\
\mbox{}\hspace{1.5cm}=\frac{2}{3}D_u^{{\Lambda}[\rm{SU}(3)]}+\frac{4}{3}
D_s^{{\Lambda}[\rm{SU}(3)]} =2 D_u^{{\Sigma^0}[\rm{SU}(3)]}=2 D_d
^{{\Sigma^0}[\rm{SU}(3)]};\\
D_d^{p[\rm{SU}(3)]}=D_u^{n[\rm{SU}(3)]}=D_s^{{\Sigma^+}[\rm{SU}(3)]}=
D_s^{{\Sigma^-}[\rm{SU}(3)]}=D_d^{{\Xi^-}[\rm{SU}(3)]}
=D_u^{{\Xi^0}[\rm{SU}(3)]}\\
\mbox{}\hspace{1.5cm}=\frac{4}{3}D_u^{{\Lambda}[\rm{SU}(3)]}-
\frac{1}{3}D_s^{{\Lambda}[\rm{SU}(3)]} =D_s
^{{\Sigma^0}[\rm{SU}(3)]},
\end{array}
\label{su3}
\end{equation}
with similar relations for the polarized fragmentation functions.
We assume that the sea quark fragmentation functions also have the
above SU(3) relations. In principle the diquark model can also be
used to partly reflect the SU(3) flavor symmetry breaking effect
if the differences in the quark, anti-diquark, and baryon masses
are taken into account in the probabilities
$a_D^{(q/B)}(z)$~\cite{a06} for a quark $q$ fragmenting into the
baryon $B$. However the SU(3) symmetry breaking effect due to this
difference in quark, diquark and baryon masses in the diquark
model is too weak to explain the experimentally measured values
for the average hadronic multiplicities per hadronic $e^+e^-$
annihilation event~\cite{PDG00}, where hyperon production is
significantly suppressed as compared with proton production.
Actually, we find that the cross sections for $\Lambda$, $\Sigma$,
and $\Xi$ baryons in $e^+e^-$ annihilation would be overestimated
by up to two orders of magnitude if we only considered this SU(3)
symmetry breaking in the framework of the diquark model. We have
to search for another possible source of the suppression effect in
hyperon production. In Ref.~\cite{Rastogi}, a description of the
strangeness suppression effect was proposed by putting a
suppression factor in the $u$, $d$, and sea quark fragmentation
functions for baryons containing a valence $s$ quark (and a
further overall suppression factor for baryons containing two $s$
quarks). In our present analysis, we will consider an alternative
suppression mechanism due to the hyperon masses which is inspired
by the string model~\cite{String}. For simplicity, we do not
include  SU(3) symmetry breaking caused by the difference in
quark, diquark and baryon masses in the diquark model itself since
this is small. We introduce an additional mass suppression factor
for the SU(3) symmetric fragmentation functions of the  diquark
model. This overall mass suppression factor should not alter
significantly the flavor  and spin structure of the fragmentation
functions as given by the diquark model. More specifically, we
assume that the quark $q$ to baryon  $B$ fragmentation function
can be expressed as follows:

\begin{equation}
D_{q}^B(z) = D_{q}^{B[\rm{SU}(3)]}(z) (2J+1) \left\{ 1 +
\frac{|S|}{(2I+1)}\right \} \exp [-b M_B^2/z^c] \label{ssp}
\end{equation}
where $S$, $I$, $J$ and $M_B$ are the strangeness, isospin,
spin and mass of the octet baryon $B$. The  term within the
curly brackets is a strangeness modification factor. The mass
suppression factor is inspired by the string model~\cite{String},
and $D_{q}^{B[SU(3)]}(z)$ is the SU(3) diquark model fragmentation
functions for the octet baryon $B$.

To summarize, our model which describes the fragmentation
functions of all the octet baryons involves a total of {\it eight}
free parameters:

\begin{equation}
N_v, \alpha_v, \beta_v, N_s, \alpha_s, \beta_s, b, c.
\end{equation}

For a fit to the experimental data, the fragmentation functions
have to be evolved from the initial scale $Q_0$ to the scale of
the experiments. We take the input scale $Q_0^2=1.0 ~\rm{GeV}^2$
and the QCD scale parameter $\Lambda_{QCD}= 0.3~ \rm{GeV}$, and
determine the free parameters of the model by fitting the
experimental data~\cite{ALEPH95,ALEPH98,OPAL97b,JPG}
on the differential cross sections

\begin{equation}
\frac{1}{\sigma_{tot}}\frac{d \sigma}{d x_E} =\frac{ \sum\limits_q
\hat{C}_q \left [ D_q^B (x_E,Q^2)+D_{\bar{q}}^B (x_E,Q^2)
\right ]} {\sum\limits_q \hat{C}_q} \label{crosection}
\end{equation}
for semi-inclusive octet baryon production  $e^+e^- \to B + X$,
where $\sigma_{tot}$ is the total cross section for the process
and $x_E=2 E_B/\sqrt{s}$. Here  $s$ is the total center-of-mass
(c.m.) energy squared, and $E_B$ the energy of the produced proton
in the $e^+e^-$ c.m. frame. In (\ref{crosection}), $\hat{C}_q$
reads

\begin{equation}
\hat{C}_q=e_q^2-2 \chi_1 v_e v_q e_q+ \chi_2 (a_e^2+v_e^2)
(a_q^2+v_q^2),\label{hatC}
\end{equation}
with
\begin{equation}
\chi_1=\frac{1}{16 \sin^2 \theta_W \cos^2 \theta_W}
\frac{s(s-M_Z^2)}{(s-M_Z^2)^2+M_Z^2\Gamma_Z^2},
\end{equation}

\begin{equation}
\chi_2=\frac{1}{256 \sin^4 \theta_W \cos^4 \theta_W}
\frac{s^2}{(s-M_Z^2)^2+M_Z^2\Gamma_Z^2},
\end{equation}
\begin{equation}
a_e=-1,
\end{equation}
\begin{equation}
v_e=-1+4 \sin^2 \theta_W,
\end{equation}
\begin{equation}
a_q=2 T_{3q},
\end{equation}
and

\begin{equation}
v_q=2 T_{3q}-4 e_q \sin^2 \theta_W,
\end{equation}
where $T_{3q}=1/2$ for $u$, while $T_{3q}=-1/2$ for $d$, $s$
quarks, $e_q$ is the charge of the quark in units of the proton
charge, $\theta$ is the angle between the outgoing quark and the
incoming electron, $\theta_W$ is the Weinberg angle, and $M_Z$ and
$\Gamma_Z$ are the mass and width of $Z^0$.

We perform a leading order (LO) analysis since the results in
Refs.~\cite{Flo98b,Kniehl00} show that the leading order fit is of
similar quality as the next-to-leading order fit.  Also, the LO
analysis should be enough in order to outline the qualitative
feature of mass suppression in baryon production. In addition, we
only use $z > 0.1$ data samples because understanding the very
low-$z$ region data needs further modifications to the evolution
of the fragmentation functions~\cite{Flo98b,Kniehl00}. However, we
find that some of data in the low-$z$ region can still be
described by our fragmentation functions. With the above mentioned
cut, we have a total of 157 experimental data
points~\cite{ALEPH95,ALEPH98,OPAL97b,JPG}. Eight free parameters
of our initial parameterizations are determined by performing a
fit to the experimental data. The total $\chi^2$ value of the fit
is 192.362, which corresponds to $\chi^2$/point=1.225. The values
of the parameters of our model are given in Table~\ref{table1}.

\begin{table*}
\caption{The parameters for the diquark model with the strangeness suppression}
\begin{tabular}{|c|c|c|c|c|c|c|c|}
\hline
%&&\\
$N_v$ & $\alpha_v$ & $\beta_v$ & $N_s$ & $\alpha_s$ & $\beta_s$
& b ($\rm{GeV}^{-2}$) & c   \\
%&&\\
\hline\raisebox{0pt}[12pt][6pt] 161.602 & 1.450 & 4.313 & 121.292 & -0.251
& 8.161 & 3.394 & 0.241  \\[4pt]
\hline
\end{tabular}
\label{table1}
%\end{footnotesize}
\end{table*}

\begin{figure*}
\includegraphics[width=10cm,height=16cm]{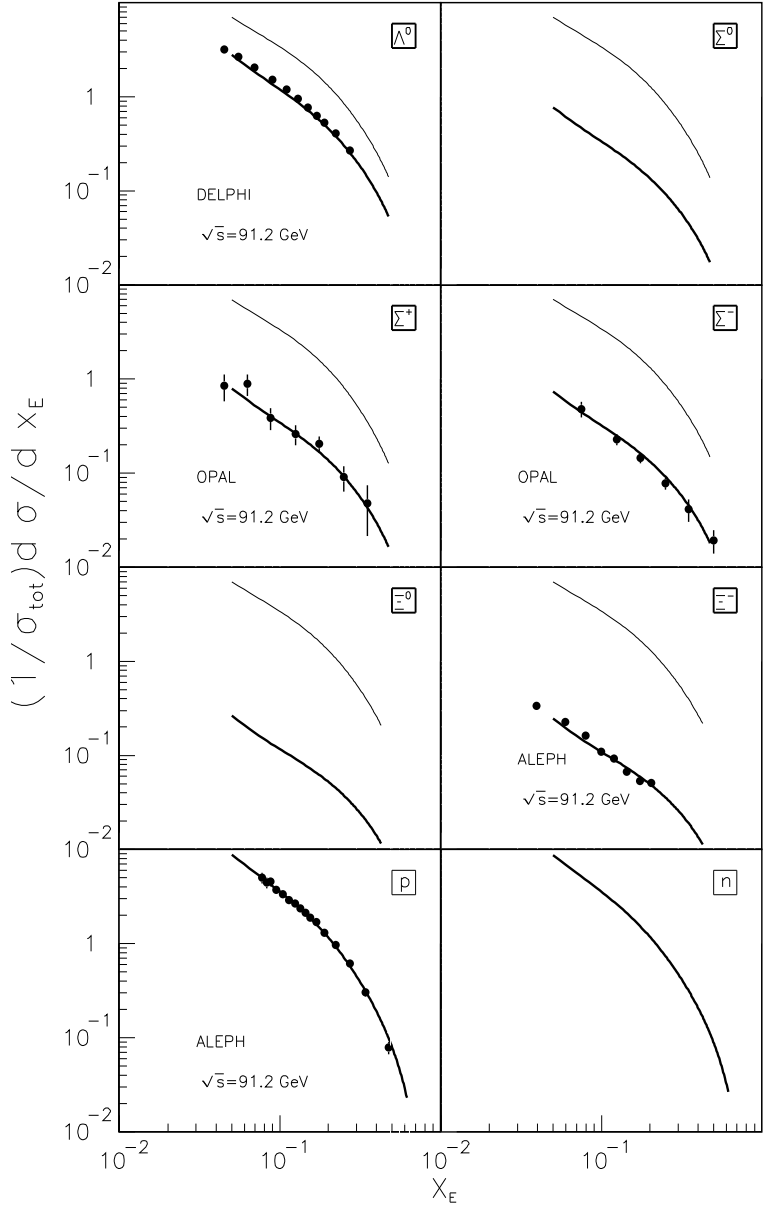}
\caption[*]{\baselineskip 13pt The comparison of our fit results
(thick solid curves) for the $x_E$ dependence of the inclusive
octet baryon production cross section $(1/\sigma_{tot})d \sigma/d
x_E$ in $e^+e^-$ annihilation and the experimental
data~\cite{ALEPH95,ALEPH98,OPAL97b,JPG}.  The thin solid curves
correspond to the results with the hyperon fragmentation functions
deduced directly from the proton fragmentation functions by using
SU(3) symmetry.} \label{ys1f1}
\end{figure*}

\begin{figure*}
\includegraphics[width=10cm,height=16cm]{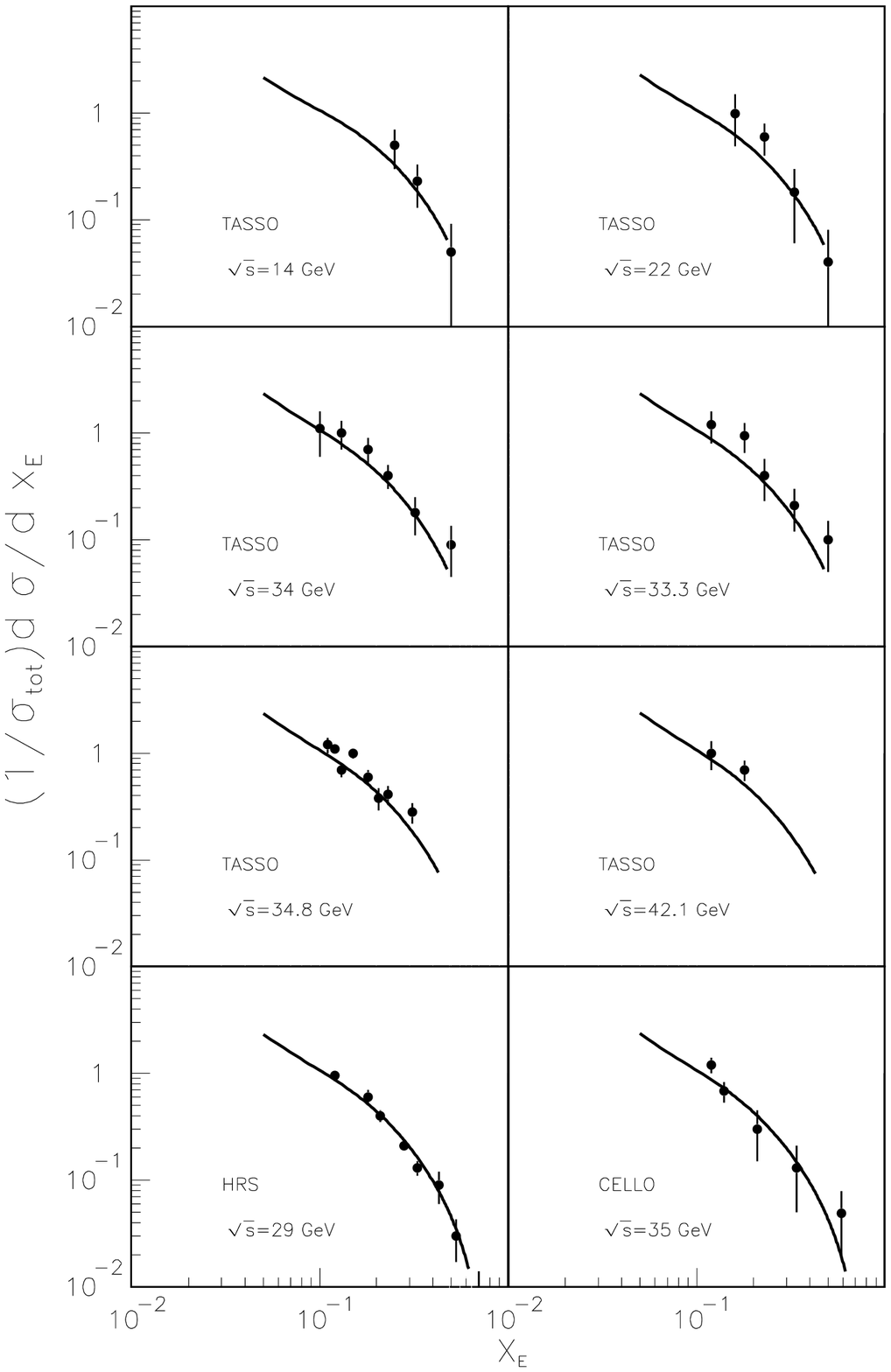}
\caption[*]{\baselineskip 13pt The comparison of our fit results
for the $x_E$ dependence of the inclusive $\Lambda$
production cross section $(1/\sigma_{tot})d \sigma/d
x_E$  in $e^+e^-$ annihilation and
the experimental data~\cite{JPG,Flo98b}.} \label{ys1f2}
\end{figure*}

In Fig.~\ref{ys1f1}, we give the fit results (thick solid curves)
as compared with the experimental data. In order to provide a
clear comparison between the experimental data and the fit curves,
only part of data are shown in the figure and a similar fit
quality is obtained for other data points. We also present the
results for the cross sections for hyperon production with the
hyperon fragmentation functions deduced directly from the proton
fragmentation functions by means of the SU(3) flavor symmetry
relation (see thin solid curves in Fig.~\ref{ys1f1}). By comparing
the thick and thin curves, one can find that the mass suppression
effect in the cross section of hyperon production is significant.
In addition, we also show in Fig.~\ref{ys1f2} the fit results for
$\Lambda$ production at various center of mass energies, which
indicates that the QCD evolution behavior of the fragmentation
functions is reasonable.

\section{Octet baryon production in charged lepton DIS}

\subsection{Unpolarized case}

In the above section, we showed a strong mass suppression effect
in hyperon production. This is an extra SU(3) symmetry breaking
distinct from that due to the quark and diquark mass differences
in the diquark model fragmentation functions, and is effectively
described by a mass suppression factor in the octet baryon
fragmentation functions. We need a cross check of this mass
suppression effect from a different process. Thus we apply the
obtained fragmentation functions to calculate the cross sections
of octet baryon production in charged lepton DIS.

\begin{figure*}
\includegraphics[width=10cm,height=16cm]{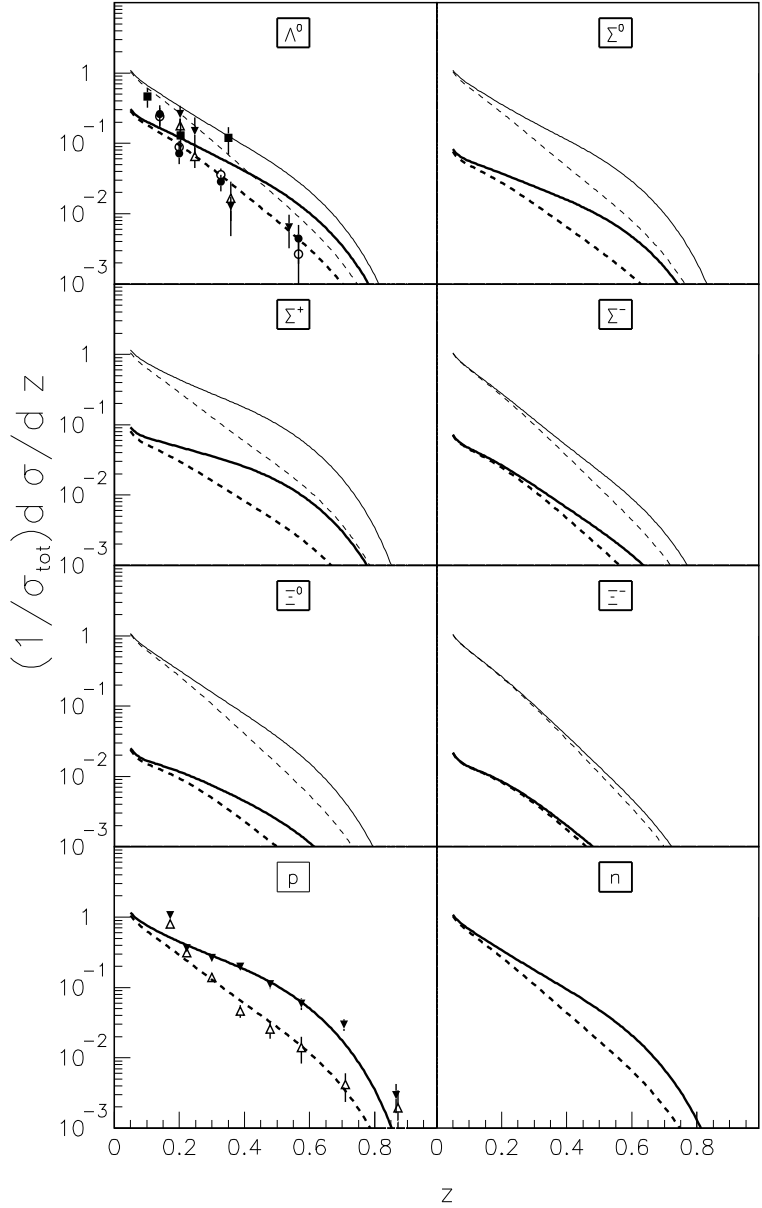}
\caption[*]{\baselineskip 13pt The cross sections for baryons
(thick solid curves) and anti-baryons (thick dashed curves)
production in charged lepton DIS, obtained with our fragmentation
functions. The results with the hyperon fragmentation functions
deduced directly from the proton fragmentation functions by using
the SU(3) symmetry are also shown for hyperons (thin solid curves)
and anti-hyperons (thin dashed curves) production. The
experimental data are taken from
Refs.~\cite{NPB480,ZPC61,PLB150}.
The original SIDIS data in terms of $x_F$
have been converted to the variable $z$ by using the method of
Ref.~\cite{Flo98b}.}\label{ys1f3}
\end{figure*}

To leading order, the cross section for the process

\begin{equation}
l + p \to B + X
\end{equation}
can be expressed as

\begin{equation}
\frac{1}{\sigma_{tot}}\frac{{\rm{d}} \sigma}{{\rm{d}} z {\rm{d}} x}
= \frac{\sum e_q^2 q(x) D_q^B(z) + (q \to
\bar{q})} {\sum e_q^2},
\end{equation}
where $q(x)$ is the quark distribution in the target nucleon. By
inserting the fragmentation functions for the octet baryons into
the above cross section, and using the CTEQ5~\cite{CTEQ5} quark
distributions in the target nucleon, we get the numerical results
shown in Fig.~\ref{ys1f3}, where the $x$-integrated cross sections
$(1/\sigma_{tot}){\rm{d}} \sigma / {\rm{d}} z$
for baryon (thick solid curves) and anti-baryon (thick dashed
curves) production are compared with the available experimental
data. In the calculation we have taken $Q^2=50~ \rm{GeV}^2$ and
the $x$ integration range [0.02, 0.4].

We find that the theoretical predictions are compatible with the
available experimental data. In Fig.~\ref{ys1f3}, we also show the
calculated results with the hyperon fragmentation functions
deduced directly from the proton fragmentation functions by using
the SU(3) symmetry (thin curves). The experimental data are taken
from Refs.~\cite{NPB480,ZPC61,PLB150}. The data points with full
circles, triangles and squares are for particle production
measured by the E665, EMC and H1 collaborations, respectively;
open circles and triangles indicate the data for anti-particle
production measured by the E665 and EMC collaborations,
respectively. The mass suppression effect in hyperon production,
especially in $\Sigma$ and $\Xi$ production, is evident.
Therefore, the hyperon production in charged DIS is an ideal place
to check the proposed mass suppression effect.

\subsection{Polarized case}

Recently experimental data on the spin transfer to $\Lambda$ in
charged lepton DIS have become available. The spin transfer is a
good observable to check the helicity structure of the
fragmentation functions for a baryon. In longitudinally polarized
charged lepton DIS on an unpolarized  proton target, the produced
baryon polarization along its own momentum axis is given in the
quark parton model by
\begin{equation}
P_{B}(x,y,z) = P_b D(y)A_{B}(x,z)~,
\label{PL}
\end{equation}
where $P_b$ is the polarization of the charged lepton beam, $D(y)$
with $y=\nu/E$ is  the longitudinal depolarization factor of the
virtual photon with respect to the parent lepton, and
\begin{equation}
A_{B}(x,z)= \frac{\sum\limits_{q} e_q^2 [q(x,Q^2) \Delta
D_q^B(z,Q^2) + ( q \rightarrow \bar q)]}
{\sum\limits_{q} e_q^2 [q (x,Q^2)
D^B_q(z,Q^2) + ( q \rightarrow \bar q)]}~,
\label{DL}
\end{equation}
is the longitudinal spin transfer to the baryon $B$.

\begin{figure*}
\includegraphics[width=10cm,height=16cm]{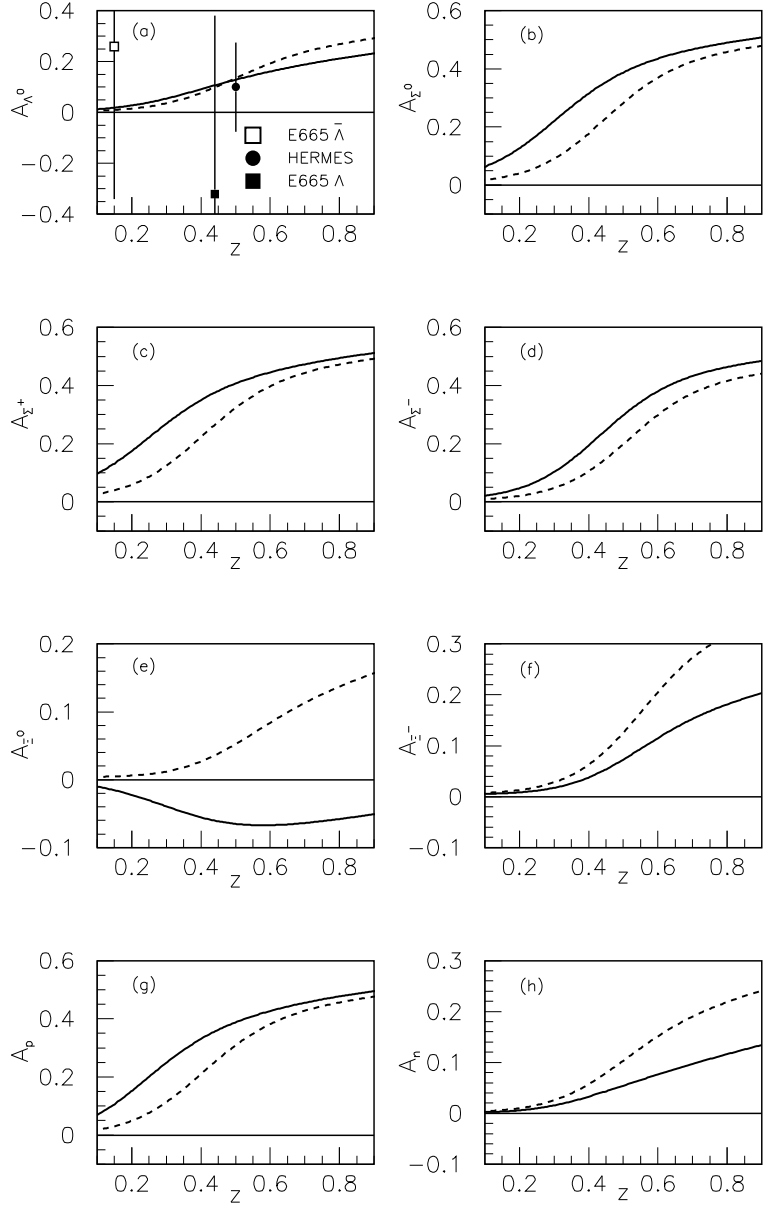}
\caption[*]{\baselineskip 13pt The spin transfer to
baryons (solid curves) and anti-baryons
(dashed curves) production
in charged lepton DIS obtained with our
fragmentation functions.
The experimental data are taken from
Refs.~\cite{HERMES,E665}.}\label{ys1f4}
\end{figure*}

In order to check the spin structure of the obtained fragmentation
functions, we calculate the $x$-integrated spin transfer to octet
baryons in charged lepton DIS. The numerical results are shown in
Fig.~\ref{ys1f4}. Our theoretical predictions are consistent with
the available experimental data on $\Lambda$ production.

\section{Octet baryon production in $pp$ collisions}

In the near future, new experimental data will become available on
hadron production in $pp$ collisions at BNL-RIHC~\cite{Saito}.
Therefore, it is interesting to predict the cross sections for octet baryon
production in $pp$ collisions in order to have a further check of
the mass suppression effect in octet hyperon production.

In leading order of perturbative QCD, the differential
cross section for the $p p \to B X$ process can be
schematically written in a factorized form as~\cite{MSSY9}

\begin{multline}
E {d^3\sigma \over d^3p}= \sum \limits_{abcd}
\int_{\bar x_a}^1 dx_a
\int_{\bar x_b}^1 dx_b f_a^{\tilde{A}}(x_a,Q^2)f_b^{\tilde{B}}(x_b,Q^2)
D_c^B(z,Q^2) {1 \over \pi z}{ d\hat \sigma \over d\hat t}(ab \to cd)~,
\label{pp1}
\end{multline}
with
\begin{equation}
\bar x_a={x_{T}e^y \over 2-x_Te^{-y}}\ ,\
\bar x_b={x_ax_Te^{-y}\over 2x_a-x_Te^y}\ ,\  z={x_T\over 2x_b} e^{-y}
+ {x_T \over 2x_a}e^y \ ,
\label{xz}
\end{equation}
where $x_T=2p_T/\sqrt s$, $\sqrt s$ is the center of mass energy
of the $pp$ collision; $p_T$, $E$ and $y$ are the transverse
momentum, energy and rapidity of the produced baryon $B$;
$f_a^{\tilde{A}} (x_a,Q^2)$ and $ f^{\tilde{B}}_b (x_b,Q^2)$ are
the unpolarized distribution functions of partons $a$ and $b$ in
the protons ${\tilde{A}}$ and ${\tilde{B}}$ at the scale
$Q^2=p_T^2$; $D_c^B( z, Q^2)$ is the fragmentation function which
we have obtained in Sec. II; $\frac{d \hat{\sigma}}{d \hat{t}}$ is
the differential cross section for the sub-process $a+b \to c +
d$, and  $\hat{t}=-x_a p_T \sqrt{s} e ^ {-y} /z$ is the Mandelstam
variable at the parton level.

By charge-conjugation invariance, the $e^+e^- \to B X$ cross
section for baryon production should be equal to that for the
corresponding antibaryon production process. Therefore, only the
combinations $D^B_q + D^{\bar{B}}_q$ can be determined, and the
same holds for the antiquark fragmentation functions. However, in
$pp$ collisions we can observe differences in the cross sections
for baryon and anti-baryon production. Therefore in this case we
also predict the cross sections for anti-baryon $\bar{B}$
production, whose quark fragmentation functions can be obtained
according to the matter-antimatter symmetry $D_{q,\bar{q}}^{B} (z)
=D _{\bar{q},q}^{\bar{B}}(z)$.

\begin{figure*}
\includegraphics[width=10cm,height=16cm]{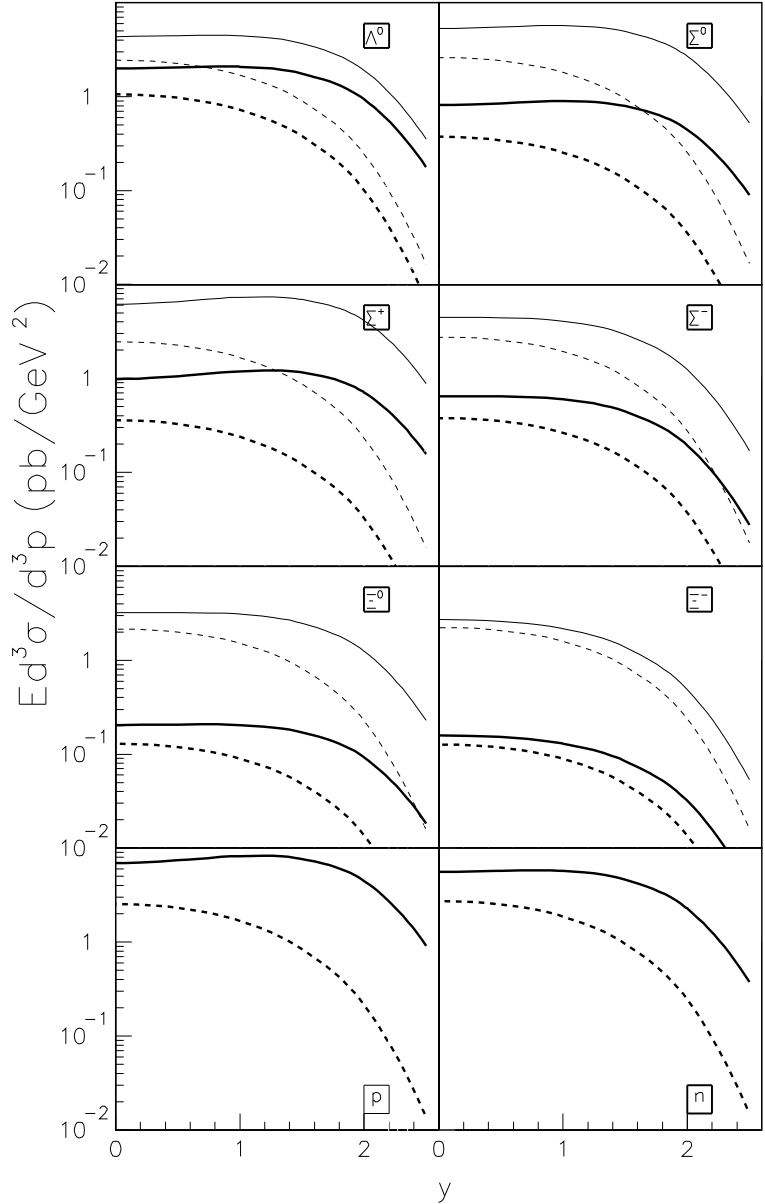}
\caption[*]{\baselineskip 13pt The cross sections for baryons
(thick solid curves) and anti-baryons (thick dashed curves)
production in $pp$ collisions are predicted at $\sqrt{s}= 500~
\rm{GeV}$ and $p_T= 15~ \rm{GeV/c}$. The results with the hyperon
fragmentation functions deduced directly from the proton
fragmentation functions by using the SU(3) symmetry relation are
also shown for hyperons (thin solid curves) and anti-hyperons
(thin dashed curves) production. }\label{ys1f5}
\end{figure*}

By adopting the LO set of unpolarized parton distributions of
Ref.~\cite{GRV95}, we present in Fig.~\ref{ys1f5} the cross
sections for octet baryons (thick solid curves) and antibaryons
(thick dashed curves) produced in $pp$ collisions. These results
are calculated at $\sqrt{s}=500~\rm{GeV}$ and $p_T=
15~\rm{GeV/c}$. As a comparison, we also calculate the cross
sections with the hyperon fragmentation functions deduced directly
from the proton fragmentation functions by using the SU(3)
symmetry relation (thin curves). By comparing the thick and thin
curves, one can find that the mass suppression effect in hyperon
production from $pp$ collisions is also significant. Therefore,
the cross sections for the octet hyperon production in $pp$
collisions should be another ideal domain where the mass
suppression effect can be checked. Although some experiments for
baryon production in $pp$ collisions have been done~\cite{pptob},
the available data were taken in the low-$p_T$ region. We need
some data at high-$p_T$ in order to check our partonic framework
predictions. This may be realized by
RHIC-BNL~\cite{Saito,MSSY9,Bunce} in the near future.

\section{Summary}

Based on the quark diquark model, the fragmentation functions for
all octet baryons are related by the SU(3) relation. Nevertheless
the hadronic multiplicities measurements in electron-positron
annihilation indicate a strong suppression in octet hyperon
production as compared with proton production, which cannot be
explained by the SU(3) symmetry breaking within the diquark model
framework. Inspired by the phenomenology of the string model, we
proposed an overall mass suppression factor for both unpolarized
and polarized octet baryon fragmentation functions, retaining the
flavor and spin structure of the fragmentation functions given by
the diquark model. We found that the diquark model with the mass
suppression factor can be used to describe quite accurately the
fragmentation functions for all octet baryons with {\it eight}
free parameters. The parameters were determined by a fit to the
available experimental data on the octet baryon production in
electron-positron annihilation. We used eight parameters; three
for the unpolarized valence down quark to proton fragmentation
function $D_{d_v}^p (z)$ (see Eq.~(\ref{fit1})), while all other
unpolarized and polarized valence quark fragmentation functions
for the proton follow from the diquark model (Eqs.~(\ref{eq16})-
(\ref{eq20})); three parameters for the sea quark fragmentation
functions (Eq.~(\ref{fit2})); and finally two more for the
suppression factor (Eq.~(\ref{ssp})). In addition, the diquark
model plays an important role in relating fragmentation functions
for all octet baryons to each other (see Eq.~(\ref{su3})).

The mass suppression factor leads to an enormous simplification in
our analysis and plays an important role in our understanding of
the experimental data on the unpolarized hyperon production in
$e^+e^-$ annihilation. This mechanism needs to be further checked.
The octet baryon fragmentation functions, determined from the
$e^+e^- \to B X$ process, can be used to predict inclusive single
baryon production cross sections in other processes, like $pp$,
$p\bar{p}$, $ep$, $\nu p$, $\mu p$ and $\gamma p$ scattering. With
the obtained fragmentation functions, we calculated the cross
section for octet baryon production in charged lepton DIS, and our
predictions are compatible with the available experimental data.
Furthermore, we predicted cross sections for octet baryon
production in $pp$ collisions. We investigated the mass
suppression effect of hyperon production in charged lepton DIS and
$pp$ collisions, and found that these two processes are ideal
places for checking the proposed mass suppression effect when
further experimental data become available.

\begin{acknowledgments}

We would like to thank A. Sch\"{a}fer and J. Soffer for valuable
discussions and comments. This work is supported by the Alexander von Humboldt
Stiftung Foundation, and partially supported by the National
Natural Science Foundation of China under Grant Number 10175074,
by the Foundation for University Key Teacher, by the Ministry of
Education (China), by the DFG and BMBF, and by Fondecyt (Chile)
project 1030355.

\end{acknowledgments}

\end{document}